\documentclass{aa}

\usepackage{txfonts}
\usepackage{astrobib}
\usepackage{graphicx}
\usepackage{marvosym}

\usepackage{longtable}
\usepackage{supertabular}

\begin{document}


  \title{The prompt to late-time multiwavelength analysis of GRB\,060210}
 
  \author{
    P.A.~Curran\inst{1}
    \and A.J.~van~der~Horst\inst{1}
    \and A.P.~Beardmore\inst{2}
    \and K.L.~Page\inst{2}
    \and E.~Rol\inst{2}
    \and A.~Melandri\inst{3}
    \and I.A.~Steele\inst{3}
    \and C.G.~Mundell\inst{3}
    \and A.~Gomboc\inst{4}
    \and P.T.~O'Brien\inst{2}
    \and D.F.~Bersier\inst{3}
    \and M.F.~Bode\inst{3}
    \and D.~Carter\inst{3}
    \and C.~Guidorzi\inst{3,5}
    \and J.E.~Hill\inst{6}
    \and C.P.~Hurkett\inst{2}
    \and S.~Kobayashi\inst{3}
    \and A.~Monfardini\inst{3,7}
    \and C.J.~Mottram\inst{3}
    \and R.J.~Smith\inst{3}
    \and R.A.M.J.~Wijers\inst{1}
    \and R.~Willingale\inst{2}
  }

  \offprints{P.A.~Curran (pcurran@science.uva.nl)}
  
  \institute{
    Astronomical Institute, University of Amsterdam, Kruislaan 403, 1098 SJ Amsterdam, Netherlands
  \and Department of Physics \& Astronomy, University of Leicester, LE1 7RH, UK
  \and Astrophysics Research Institute, Liverpool John Moores University, Twelve Quays House, Birkenhead, CH41 1LD, UK
  \and Faculty of Mathematics and Physics, University of Ljubljana, Jadranska 19, 1000 Ljubljana, Slovenia
  \and Present Address: INAF - Osservatorio Astronomico di Brera, via Bianchi 46, 23807 Merate (LC), Italy
  \and NASA Goddard Space Flight Center, Greenbelt, MD 20771, USA
  \and Present Address: Institut N\'{e}el, CNRS, 25 Avenue des Martyrs, 38000 Genoble, France
  }

\date{Received ; accepted}

\abstract {} 
{We present our analysis of the multiwavelength photometric \& spectroscopic observations of GRB\,060210 and discuss the results in the overall context of current GRB models.}
{All available optical data underwent a simultaneous temporal fit, while X-ray and $\gamma$-ray observations were analysed temporally \& spectrally. The results were compared to each other and to possible GRB models.}
{The X-ray afterglow is best described by a smoothly broken power-law with a break at 7.4 hours.
The late optical afterglow has a well constrained single power-law index which has a value between the two X-ray indices, though it does agree with a single power-law fit to the X-ray.
An evolution of the hardness of the high-energy emission is demonstrated and we imply a minimum host extinction from a comparison of the extrapolated X-ray flux to that measured in the optical.
}
{We find that the flaring $\gamma$-ray and X-ray emission is likely due to internal shocks while the flat optical light curve at that time is due to the external shock. 
The late afterglow is best explained by a cooling break between the optical and X-rays and continued central engine activity up to the time of the break. The required collimation corrected energy of $\sim 2 \times 10^{52}\,\mathrm{erg}$, while at the high end of the known energy distribution, is not unprecedented. 
}


\maketitle


\section{Introduction} \label{intro}

Since the launch of the \emph{Swift} satellite \cite{2004ApJ...611.1005G} we have been allowed observations of Gamma Ray Bursts (GRBs) at very early times, occasionally observing in X-rays and optical light even as the prompt $\gamma$-ray emission is ongoing. 
Gradually, evidence has emerged for a fairly canonical behaviour of $\gamma$-ray and X-ray light curves
in which the X-rays decay fairly rapidly after the prompt emission and then decay more slowly for a while before becoming steeper again \cite{2006ApJ...642..389N}. 
This behaviour is seen in many of the bursts, but a significant minority decay more gradually from early times. These differences may be due to the different relative strengths of emission components due to internal and external processes \cite{2006ApJ...647.1213O}.
Similar complexity is seen in the optical, where some bursts show optical behaviour which does not exactly mimic the X-ray, again suggesting a contribution from several emission components. 
Here we present multiwavelength observations of GRB\,060210, covering the early to late-time emission and compare the data to the expected canonical behaviour and the blast wave afterglow model (e.g. \citeNP{2004IJMPA..19.2385Z}).

GRB\,060210 is a long burst, detected by the Burst Alert Telescope (BAT) on board \emph{Swift} \cite{2005SSRv..120..143B} on February 10th, 2006 at 04:58:50 UT. 
The X-Ray Telescope (XRT; \citeNP{2005SSRv..120..165B}), which started observing the region 95 seconds after the BAT trigger, identified the X-ray counterpart position to within an error of 5.4$^{\prime\prime}$ \cite{2006GCN..4724....1B}.
While the Ultraviolet/Optical Telescope (UVOT; \citeNP{2005SSRv..120...95R}) was unable to identify an optical counterpart, the robotic Palomar 60'' telescope observed the R-band counterpart at approximately 5.5\,min post trigger \cite{2006GCN..4723....1F}. The redshift was estimated as $z = 3.91$ \cite{2006GCN..4729....1C} and a  host galaxy has been proposed based on near infrared observations \cite{2006GCN..4753....1H}. No radio emission has been observed down to a $3\sigma$ limit of 72\,$\mu$Jy at 8.64\,GHz
\cite{2006GCN..4761....1F}.

In this paper we present our multi-wavelength analysis of the burst from $\gamma$-ray, through X-ray to optical wavelengths. Throughout, we use the convention that a power-law flux is given as $F_{\nu} \propto t^{-\alpha} \nu^{-\beta}$ where $\alpha$ is the temporal decay index and $\beta$ is the spectral index.
The power-law of spectra is given as $\mathrm{d}N / \mathrm{d}E \propto E^{-\Gamma}$ where $\Gamma$ is the photon index and is related to spectral index as $\Gamma = \beta + 1$.
In \S\ref{obs} we introduce our observations and reduction methods of each wavelength regime. In \S\ref{results} we present the results of our spectral and temporal analysis, while in \S\ref{dis} we discuss these results in the overall context of the burst.


\section{Observations}\label{obs}

\subsection{$\gamma$-ray} \label{obs-bat}

The \emph{Swift}-BAT triggered and localised GRB\,060210 at $T_{0} =$ 04:58:50 UT and promptly distributed the coordinates via the Gamma-ray burst Coordinates Network (GCN). Data for the burst were obtained in event mode, covering $T_0 - 300$\,s to $T_0 + 300$\,s, giving spectral coverage from 15 -- 350\,keV and time resolution of 64 ms. 

\begin{figure} 
  \centering 
  \resizebox{\hsize}{!}{\includegraphics[angle=-90]{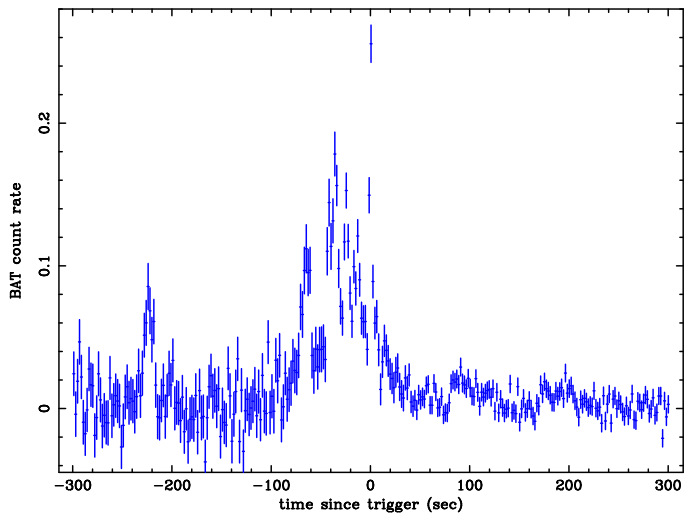}}
  \caption{Full BAT light curve (15 -- 350\,keV) of GRB\,060210, clearly showing the main peak and post-trigger activity, as well as the pre-trigger activity at $\sim T_{0} - 230$\,s.} 
  \label{bat.lc} 
\end{figure}

BAT event data were reduced and analysed within FTOOLS v6.0.4. Light Curves and spectra were extracted by the standard methods, and standard corrections applied \cite{2006ApJ...647.1213O}.
From the light curve (figure \ref{bat.lc}) the main period of activity is identified as being from $T_0 - 75$\,s to $T_0 + 20$\,s with the brightest peak at $T_0$. However significant emission was recorded from $T_0 - 230$\,s to the last peak at $T_0 + 200$\,s, which represents a significant fraction of the total recorded event data.  The $T_{90}$ in the BAT range (15-350\,keV) is calculated as $220 \pm 70$\,s, covering the period from $\sim T_0 - 149$\,s to $\sim T_0 + 70$\,s.

\subsection{X-ray} \label{obs-xrt}

The XRT started observing at 05:00:25 UT, 95 seconds after the initial BAT trigger, and found an uncatalogued, variable X-ray source which was identified as the afterglow. XRT 
observed up to March 3, or approximately $2 \times 10^6$ seconds post-burst, by which time only an upper limit detection was obtainable from the data. Data were initially collected in Windowed Timing (WT) mode
\cite{2004SPIE.5165..217H,2005SPIE.5898..313H}, which gives no positional information but high temporal resolution. After the initial $\sim500$\,s of WT mode, data were collected in Photon Counting (PC) mode.

The XRT data were initially processed with the FTOOL, \texttt{xrtpipeline} (v0.9.9) with \texttt{wtbiasdiff} applied to correct for potential WT bias-row subtraction problems, caused by bright Earth or CCD temperature variations. 
Background subtracted 0.3-10keV light curves were extracted from the cleaned event lists with a minimum of 20 counts/bin. Likewise, source and background spectra were extracted for analysis with \texttt{Xspec}.
The first orbit of PC mode data were corrected for pile-up as detailed in \citeN{2006ApJ...638..920V}. 
A correction was also made, using \texttt{xrtexpomap}, for fractional exposure loss due to bad columns on the CCD, which arose after damage caused by a micrometeroid strike \cite{2006xru..conf..943A}.

\begin{figure} 
  \centering 
  \resizebox{\hsize}{!}{\includegraphics[angle=-90]{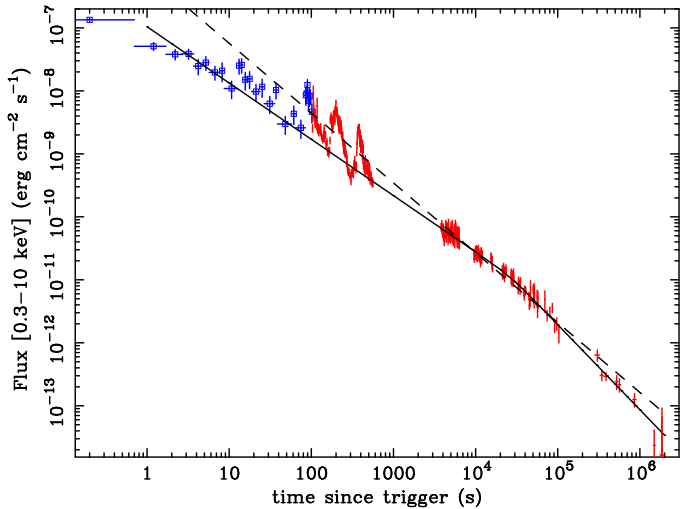} }
  \caption{High-energy light curve of GRB\,060210 composed of the XRT data (red crosses) and the BAT data (blue boxes) extrapolated to the XRT energy range. The dashed line shows a single power-law fit to the late afterglow with $\alpha = 1.11 \pm 0.02$, but the solid line showing a smoothly broken power-law with indices $\alpha_{1} = 0.89 \pm 0.05$ \& $\alpha_{2} = 1.35 \pm 0.06$ is a significantly better fit.} 
  \label{bat-xrt.lc} 
\end{figure}

Spectral fits are used to convert count rates of both WT \& PC mode data to fluxes, and to extrapolate the BAT light curve to the XRT energy range (\citeNP{2006ApJ...647.1213O}; figure \ref{bat-xrt.lc}). 
Initial analysis of the early time part of this light curve shows two strong flares at $\sim$200\,s and $\sim$380\,s, after an initial decay that seems to be a continuation of a flare at $\sim$100\,s as observed in $\gamma$-rays.

\subsection{Optical} \label{obs-opt}

Optical observations in B, R and i$^\prime$ bands were obtained by the robotic, 2.0\,m  Faulkes Telescope North (FTN) at Haleakala on Maui, Hawaii; FTN is the sister instrument of The Liverpool Telescope \cite{2006PASP..118..288G}. 
At 05:04:36 UT, approximately 5.8\,min post-burst, the telescope triggered its automatic 1\,hour-sequence and to ensure good time coverage of the light curve, a second 1\,hour sequence was manually triggered at 06:05:27 UT, 66\,min post-burst. The previously reported optical transient \cite{2006GCN..4723....1F} was clearly visible in R and i$^\prime$ but not in B (Table \ref{optical_log}). This is not unexpected, as the high redshift of the burst and the associated extinction due the Lyman forest causes significant dimming in the B-band while i$^\prime$ remains unaffected.

\begin{table}	
  \begin{center} 	
    \caption{FTN optical observations of GRB\,060210 calibrated to the Cousins photometric system, uncorrected for Galactic extinction of $E_{(B-V)} = 0.093$. Magnitude errors are at $1\sigma$ level.} 	
    \label{optical_log} 	
    \begin{tabular}{l l l l} 
      $T_{\mathrm{mid}}-T_0$  & $T_{\mathrm{exp}}$ & Band & Mag\\ 
      (min)  & (min) &  &  \\
      \hline \hline
      65.58   & 31.33   & B  &  22.10$^{a}$  \\
      \hline
      5.85    & 0.1667  & R   &	18.50  $\pm$  0.20 \\ 
      6.20    & 0.1667  & R   &	18.36  $\pm$  0.15 \\ 
      6.55    & 0.1667  & R   &	18.36  $\pm$  0.15 \\ 
      11.85   & 0.5     & R   &	18.51  $\pm$  0.12 \\ 
      16.35   & 1.0     & R   &	18.87  $\pm$  0.10 \\ 
      23.20   & 2.0     & R   &	19.43  $\pm$  0.10 \\ 
      33.04   & 3.0     & R   &	19.74  $\pm$  0.10 \\ 
      43.25   & 2.0     & R   &	20.14  $\pm$  0.15 \\ 
      53.18   & 3.0     & R   &	19.98  $\pm$  0.15 \\  
      72.00   & 2.0     & R   & 20.57  $\pm$  0.20 \\
      88.84   & 5.0     & R   &	20.8   $\pm$  0.4  \\
      115.26  & 9.0     & R   &	21.2   $\pm$  0.3  \\
      167.0   & 30.0    & R   &	22.4   $\pm$  0.5  \\
      \hline
      9.43    & 0.1667  & I  &	16.94  $\pm$  0.08 \\ 
      13.19   & 0.5     & I  &	17.20  $\pm$  0.05 \\ 
      18.11   & 1.0     & I  &	17.66  $\pm$  0.05 \\ 
      25.95   & 2.0     & I  &	18.25  $\pm$  0.06 \\ 
      36.98   & 3.0     & I  &	18.56  $\pm$  0.06 \\ 
      46.00   & 2.0     & I  &	18.80  $\pm$  0.08 \\ 
      57.10   & 3.0     & I  &	19.18  $\pm$  0.08 \\  
      70.12   & 0.1667  & I  &  19.35  $\pm$  0.25 \\
      73.73   & 0.5     & I  &	19.43  $\pm$  0.20 \\
      78.73   & 1.0     & I  &	19.55  $\pm$  0.20 \\
      86.41   & 2.0     & I  &	19.85  $\pm$  0.30 \\
      97.26   & 3.0     & I  &	20.0   $\pm$  0.3  \\
      106.13  & 2.0     & I  &	20.2   $\pm$  0.3  \\
      116.97  & 3.0     & I  &	20.3   $\pm$  0.3  \\
      \hline
    \end{tabular}
\begin{list}{}{}
\item[$^{\mathrm{a}}$] $3\sigma$ upper limit
\end{list}

  \end{center}
\end{table}

The data were analysed using the IRAF package\footnote{IRAF is distributed by the National Optical Astronomy Observatories, which are operated by the Association of Universities for Research in Astronomy, Inc., under cooperative agreement with the National Science Foundation.}, wherein reduction \& differential photometry were carried out.  
Landolt photometric standards taken on the night were used  to calibrate the magnitude of the optical counterpart. 
The i$^\prime$ magnitudes were converted to the corresponding I-band magnitudes using the observed filter transformations of \citeN{2002AJ....123.2121S}, which for normal colours are in agreement with those of \citeN{1996AJ....111.1748F}. 
The magnitudes are uncorrected for Galactic extinction of $E_{(B-V)} = 0.093$ \cite{1998ApJ...500..525S}. Astrometric calibration was carried out with respect to the NOMAD catalogue\footnote{http://www.nofs.navy.mil/nomad/}
to refine the position of the burst to within 0.2$^{\prime\prime}$ to be 03:50:57.36, +27:01:34.4 (J2000).

\section{Results} \label{results}

We have split our discussion of the emission into two phases, covering what we refer to as the \emph{early emission} and the \emph{late afterglow}. We take the early emission as being the $\gamma$-ray and the flaring X-ray emission at times $\lesssim 600$ seconds, which follows from the X-ray light curve (figure \ref{bat-xrt.lc}). We also include up to $\sim 600$ seconds of optical observations, those not displaying a power-law decay, as being early emission. The late afterglow, we take as the smooth X-ray and optical light curves, obeying power-law decay. All uncertainties of light curve analysis, spectral fits and other are quoted at the 90\% confidence level.

\subsection{Early high-energy emission} \label{results_early}

The prompt emission as measured by BAT shows a main peak which decays smoothly until $\sim$10\,s (figure \ref{bat-xrt.lc}). From $\sim$10\,s to $\sim$600\,s, GRB\,060210 is observed to display flaring in both the $\gamma$-rays and X-rays. One flare, at $\sim$100\,s, is observed by both BAT \& XRT. This flaring behaviour seems to be superimposed on an underlying power-law decay due to the afterglow. 
The time-averaged BAT spectrum (15 -- 150\,keV) over $T_{90}$ 
was fit with an unabsorbed power-law in \texttt{Xspec} and a photon index, $\Gamma = 1.55 \pm 0.09$ was found ($\chi^{2}_{\nu} = 0.84$, 56 d.o.f.). This spectral fit corresponds to a fluence of $(6.0^{+0.1}_{-0.7})\times10^{-6}$\,erg cm$^{-2}$.

\begin{figure} 
  \centering 
   \resizebox{\hsize}{!}{\includegraphics[angle=-90]{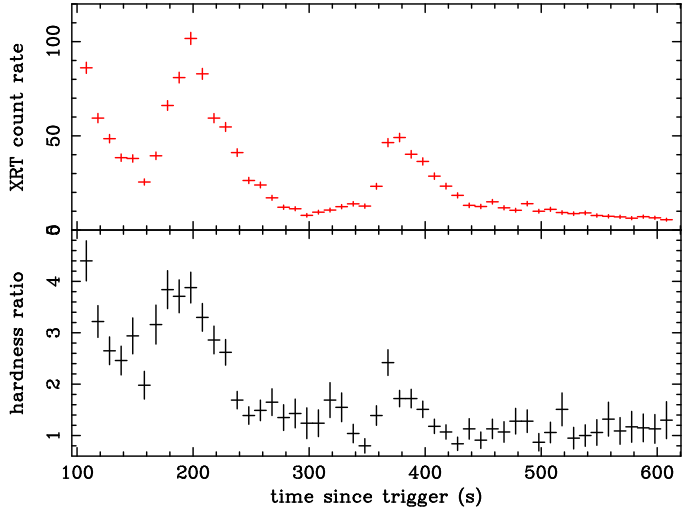} }
  \caption{\emph{(top)} Light Curve of the Windowed Timing mode data of GRB\,060210 and \emph{(bottom)} the hardness ratio (1-10\,keV/0.3-1\,keV) plot which clearly displays an evolution coupled to the count rate.} 
  \label{hardness} 
\end{figure}

When the XRT WT-mode spectrum, from 103\,s -- 614\,s is fit with an absorbed power-law, we find a photon index of  $\Gamma = 2.09 \pm 0.04$ and a column density of $N_{\mathrm{H}} = (16 \pm 1) \times 10^{20}$\,cm$^{-2}$  ($\chi^{2}_{\nu} = 1.020$, 316 d.o.f.). This is significantly higher than the Galactic value of $N_{\mathrm{H}} = 8.5 \times 10^{20}$\,cm$^{-2}$ \cite{1990ARA&A..28..215D}. This excess extinction may be explained either by rest frame extinction or by a broken power-law. Fitting an absorbed broken power-law gives a column density consistent with that of the Galactic value, so we fix the parameter to be the Galactic value. This leads to a photon index of $\Gamma = 1.96 \pm 0.03$ above a break energy, $E_{\mathrm{break}} = 0.71 \pm 0.07$\,keV and an index of $\Gamma = 0.6 \pm 0.3$ ($\chi^{2}_{\nu} = 0.923$, 315 d.o.f.) below the break. Alternatively, assuming solar abundances, a rest frame column density of $N_{\mathrm{H}} = (2.3 \pm 0.3) \times 10^{22}$\,cm$^{-2}$ in combination with the Galactic value gives an equally good fit ($\chi^{2}_{\nu} = 0.936$, 316 d.o.f.) and implies a photon index, $\Gamma = 2.03 \pm 0.03$. 
Assuming LMC and SMC abundances \cite{1992ApJ...395..130P} in the host implies $N_{\mathrm{H}} = (5.7 \pm 0.7) \times 10^{22}$\,cm$^{-2}$ ($\chi^{2}_{\nu} = 0.948$, 316 d.o.f.) and $N_{\mathrm{H}} = (10.3 \pm 1.2) \times 10^{22}$\,cm$^{-2}$ ($\chi^{2}_{\nu} = 0.950$, 316 d.o.f.) respectively, while leaving the photon index unchanged within uncertainties.Though it is not possible to favour one model over the other, we see that the photon index, $\Gamma \sim 2.0$ is marginally different to that of the earlier $\gamma$-ray spectrum. 
A hardness ratio (1-10\,keV/0.3-1\,keV) plot of the X-ray emission (figure \ref{hardness}) confirms spectral evolution, though the $\gamma$-rays are too faint to obtain a similar plot. This evolution is similar to that observed in the $\gamma$-rays for a number of BATSE bursts \cite{1994ApJ...426..604B}. 

\begin{table}[ht]	
  \begin{center} 	
    \caption{Rest frame column densities as found in combination with Galactic extinction. } 	
    \label{host_NH} 	
    \begin{tabular}{l c} 
      & $N_{\mathrm{H}}$ \\
                       & ($\times 10^{22}$\,cm$^{-2}$) \\
      \hline \hline
      SMC            & 10.3 $\pm$ 1.2 \\
      LMC            & 5.7 $\pm$ 0.7 \\
      Galactic       & 2.3 $\pm$ 0.3 \\
      \hline
    \end{tabular}
  \end{center}
\end{table}

Using the average photon index of the BAT \& XRT spectra, we get a conversion from BAT count rate to unabsorbed flux in the 0.3 -- 10\,keV range of $4.38 \times 10^{-7}$\,erg\,cm$^{-2}$\,cts$^{-1}$. Using the same fit of the XRT spectrum, yields a conversion of $4.83 \times 10^{-11}$\,erg\,cm$^{-2}$\,cts$^{-1}$ which we apply to the WT mode data.

\subsection{Late X-ray afterglow} \label{results_late}

The late X-ray afterglow was observed from $\sim$1\,hour to $\sim$23 days, at which stage only an upper limit determination was possible. The light curve for this period is quite smooth, with no obvious signs of flares or bumps, and a hardness ratio plot of the afterglow shows no signs of evolution. 
The spectrum gives results comparable with the prompt spectrum (\S \ref{results_early}), including an excess extinction over that of the Galactic value.
A broken power-law with Galactic absorption suggests a photon index of $\Gamma = 2.13 \pm 0.06$ above a break at  $E_{\mathrm{break}} = 1.04 \pm 0.15$\,keV and an index of $\Gamma = 1.4 \pm 0.2$ ($\chi^{2}_{\nu} = 1.021$, 225 d.o.f.) below the break. 
Assuming a single power-law with Galactic absorption, the rest frame absorption assuming solar abundances is consistent with that found during the prompt emission and hence is fixed at that value. 
The photon index is then $\Gamma = 2.14 \pm 0.03$ ($\chi^{2}_{\nu} = 1.015$, 227 d.o.f.). 
It is not possible to favour one model over the other but we can assume a corresponding spectral index, $\beta = 1.14 \pm 0.03$, at least at energies $\gtrsim 1.0$\,keV. Using the single power-law fit with rest frame absorption yields a PC mode, count rate to flux conversion of $5.12 \times 10^{-11}$\,erg\,cm$^{-2}$\,cts$^{-1}$.

Fitting a power-law to the temporal decay we find 
$\alpha = 1.11 \pm 0.02$ ($\chi^{2}_{\nu} = 1.69$, 144 d.o.f.),  
though an F-Test shows that a smoothly broken power-law with 
$\alpha_{1} = 0.89 \pm 0.05$,  $\alpha_{2} = 1.35 \pm 0.06$ and a break at 
$t_{\mathrm{break}} = 7.4 ^{+2.1}_{-1.6}$\,hr ($\chi^{2}_{\nu} = 0.954$, 142 d.o.f.) is a significantly better fit, in agreement with the analysis of \citeN{2006GCN..5147....1D}. 
Extrapolating the power-law fit to early times, we see that it does not match the early emission, which implies a break at $\sim 10^{3}$\,s. Doing likewise for the broken power-law fit, we see that it matches quite well with the troughs of the flares and the extrapolated BAT emission (figure \ref{bat-xrt.lc}). 
To investigate any possible spectral difference, we split the spectrum into pre- and post-break sections but the resultant fits were consistent with each other and the combined fit. This implies that any difference in spectral indices must be smaller than the errors, $\delta\beta \lesssim 0.05$.

\subsection{Optical emission} \label{results_optical}

Fitting our optical light curves of the late afterglow ($T \gtrsim 500$\,s) with a single power-law, we  find 
$\alpha_{\mathrm{R}} = 1.07 \pm 0.11$  ($\chi^{2}_{\nu} = 1.07$) and 
$\alpha_{\mathrm{I}} = 1.19 \pm 0.05$ ($\chi^{2}_{\nu} = 0.86$).
To make full use of the available data we simultaneously fit our data and previously publised data, at times $> 540$\,s, over 5 bands: FTN R, FTN I, MDM R, KAIT I, \& KAIT unfiltered (\citeNP{2006GCN..4727....1L,2007ApJ...654L..21S}). 
To do this we assume that each has the same temporal decay but we make no assumptions regarding the relative offsets -- thus eliminating any error in absolute magnitude estimates. This fit uses the \emph{simulated annealing} method ($\S$ 10.9 of \citeNP{1992nrca.book.....P}, and references therein) 
to minimise the combined $\chi^{2}$ of the data sets and hence find the best fit parameters, while a Monte Carlo analysis with synthetic data sets is carried out to estimate the errors. From this we find an optical temporal decay, 
$\alpha_{\mathrm{opt}} = 1.15 \pm 0.04$  ($\chi^{2}_{\nu} = 1.12$, 38 d.o.f.)
and magnitude offsets, which are the model magnitudes evaluated at 1 second, as detailed in table \ref{optical_fit}. 
This decay is shallower than the $\alpha \sim 1.3$ found by \citeN{2007ApJ...654L..21S}.

\begin{table}[ht]	
  \begin{center} 	
    \caption{Simultaneous temporal fit of optical data, where Mag is the model magnitude evaluated at 1 second. The common temporal decay index, $\alpha_{\mathrm{opt}}$ is fit as $1.15 \pm 0.04$.} 	
    \label{optical_fit} 	
    \begin{tabular}{l l} 
      Band     & Mag \\
      \hline 
      FTN~R            & 10.22 $\pm$ 0.34 \\
      FTN~I            & 8.92  $\pm$ 0.30 \\
      KAIT~I           & 9.28  $\pm$ 0.30 \\
      KAIT~unfiltered  & 9.81  $\pm$ 0.30 \\
      MDM~R            & 10.11 $\pm$ 0.34 \\
      \hline
    \end{tabular}
  \end{center}
\end{table}

\begin{figure} 
  \centering 
  \resizebox{\hsize}{!}{\includegraphics[angle=-90]{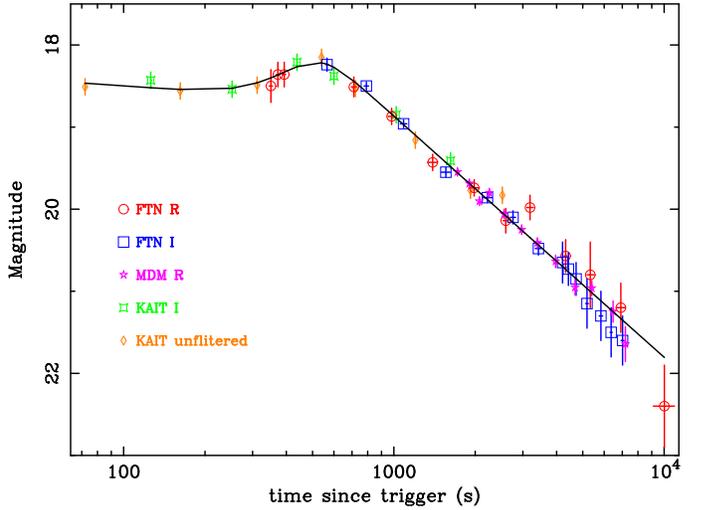} }
  \caption{Faulkes Telescope North  R and I band light curves, supplemented by previously published data, shifted to the FTN R-band offset. Fit shown is a 3-index smoothly broken power-law.}
  \label{optical.lc} 
\end{figure}

Shifting all the data to the FTN R magnitude offset confirms that there is no structure above noise in the optical light curve (figure \ref{optical.lc}). It is possible that the light curve is breaking at $\sim5000$ seconds, but there are not enough data to support this. Fitting a smoothly broken power-law with three indices to this shifted data, we find that the two break times are at 
$310 \pm 8$\,s and $540 \pm 6$\,s and the indices are 
$\alpha_{1} = 0.10 \pm  0.10$, $\alpha_{2} = -0.90 \pm 0.35$, $\alpha_{3} = 1.175 \pm 0.016$ 
($\chi^{2}_{\nu} = 0.89$, 47 d.o.f.).

\section{Discussion} \label{dis}

It has been claimed \cite{2007ApJ...654L..21S} that this burst closely resembles \object{GRB\,050801} \cite{2006ApJ...638L...5R} which is true insofar as both display an initially flat optical light curve. 
However GRB\,060210  displays a peak before decaying as a power law, while GRB\,050801 shows no such peak. 
Furthermore, the X-ray behaviour of GRB\,050801 seems to be consistent with the optical, but in GRB\,060210 they seem to be unconnected.
This suggests that in GRB\,050801 the X-ray and optical emission is originating in the same region, while this is not true for GRB\,060210. 
In the case of GRB\,060210 the early flaring X-ray \& $\gamma$-ray emission implies internal shocks, indicative of prolonged engine activity \cite{2005ApJ...630L.113K}. The optical emission, on the other hand, is consistent with an external shock afterglow.

\subsection{X-ray flares} \label{diss_flares}

The spectral evolution exhibited in the hardness ratio plot (figure \ref{hardness}) supports internal shocks as the source of the flaring X-ray emission as suggested for \object{GRB\,050607} by \citeN{2006ApJ...645.1315P}.  
Though in this case the spectral index of the flaring period is similar to that of the late afterglow, the afterglow is softer, following the trend exhibited by the hardness ratio plot. It should be noted also, that the spectral index of the flaring period is an average over that period. The difference between the two values and the lack of spectral evolution in the late afterglow certainly suggests two distinct origins for energy emission. Combined with the overlap of the last $\gamma$-ray and first X-ray flare, it seems likely that these two would share internal shocks as the common emission mechanism.

\citeN{2006ApJ...646..351L} 
show that X-ray flares can be modeled  by the curvature effect \cite{2000ApJ...541L..51K}
which causes a decay of  $F \propto (t - t_{\mathrm{ej}})^{-(2+\beta)}$ superimposed on the regular afterglow decay of  $F \propto t^{-\alpha}$, where $t_{\mathrm{ej}}$ is the energy injection time of that flare. We tested the flaring X-ray data against the curvature effect, using the average spectral index over that period and find ejection times for the 3 main X-ray flares at $\sim$ 100\,s, 200\,s \& 380\,s of $72 \pm 6$\,s,
 $155 \pm 16$\,s 
\& $310 \pm 16$\,s 
respectively. These are consistent with \citeANP{2006ApJ...646..351L} -- 
insofar as the energy ejection times are at the the start of the rising segment of the flare -- supporting the claim that the flares are indeed due to internal shocks.

\subsection{Spectral indices \& host extinction} \label{diss_spec}

\begin{table*}[ht]	
  \begin{center} 	
    \caption{Host extinction and column density to extinction ratio assuming SMC, LMC \& Galactic extinction curves in the host as discussed in \S \ref{diss_spec}. In the case where $\nu< \nu_{c}$, $\nu_{c} = 10^{17}$\,Hz.} 	
    \label{extinction} 	
\begin{tabular}{c c c c c c c c c c c c}

    & & & &  SMC & & & LMC & & & Galactic &  \\
  \hline
    & Band &  A$_{\nu(1+z)}$ & & E$_{\mathrm{B-V}}$ & $N_{\mathrm{H}}$/E$_{\mathrm{B-V}}$  & & E$_{\mathrm{B-V}}$ & $N_{\mathrm{H}}$/E$_{\mathrm{B-V}}$ & & E$_{\mathrm{B-V}}$  & $N_{\mathrm{H}}$/E$_{\mathrm{B-V}}$ \\

 & & & &  & $\times 10^{22}$\,cm$^{-2}$ & & & $\times 10^{22}$\,cm$^{-2}$ & &  & $\times 10^{22}$\,cm$^{-2}$ \\

      \hline \hline
      $\nu_{c} < \nu$  & R   & 6.7 $\pm$ 0.6 & &   0.45 $\pm$ 0.04 & 23 $\pm$ 5  & &   0.55 $\pm$ 0.05  & 10 $\pm$ 2 & &  0.78 $\pm$ 0.08 &  2.9 $\pm$ 0.7 \\
                       & I   & 6.1 $\pm$ 0.6 & &   0.57 $\pm$ 0.06 & 18 $\pm$ 4  & &  0.65 $\pm$ 0.06  & 9  $\pm$ 2 & &  0.81 $\pm$ 0.08 &  2.8 $\pm$ 0.7  \\
      \hline
      $\nu< \nu_{c}$   & R   & 3.9 $\pm$ 0.7 & &   0.26 $\pm$ 0.04 & 40 $\pm$ 10 & &  0.32 $\pm$ 0.05  & 18 $\pm$ 5 & &  0.45 $\pm$ 0.08 &  5.0  $\pm$ 1.6 \\
                       & I   & 3.0 $\pm$ 0.7 & &   0.28 $\pm$ 0.06 & 37 $\pm$ 12 & &  0.32 $\pm$ 0.07  & 18 $\pm$ 6 & &  0.40 $\pm$ 0.09 &  5.8  $\pm$ 2.0  \\

      \hline
    \end{tabular}
\end{center}
\end{table*}

From the fitted flux of the FTN optical data points at 5000\,s, corrected for galactic extinction and extinction due to Lyman absorption \cite{1995ApJ...441...18M},  we find an optical spectral index of $\beta_{\mathrm{opt}} = 3.1 \pm 0.4$. 
Converting the X-ray flux at the same time to mJy using the X-ray spectral index $\beta_{\mathrm{X}}= 1.14$, gives a flux $F_{1.732\,\mathrm{keV}} = 3.27 \times 10^{-3}$\,mJy. 
These fluxes correspond to an optical to X-ray spectral index, $\beta_{\mathrm{opt-X}} = 0.3 \pm 0.1$, confirming the optical to X-ray flux ratio of $F_{\nu\mathrm{,opt}}/F_{\nu\mathrm{,X}} \sim 10$ of \citeANP{2007ApJ...654L..21S}.

The large differences between  $\beta_{\mathrm{opt}}$, $\beta_{\mathrm{X}}$ and $\beta_{\mathrm{opt-X}}$ imply that there could be an amount of host extinction that we have failed to take into account. To estimate this host extinction, we extrapolate the X-ray flux to optical magnitudes to measure the optical extinction, A$_{\nu(1+z)}$ above that of Galactic.
From this we can calculate the corresponding value of E$_{\mathrm{B-V}}$ assuming SMC, LMC and Galactic extinction curves \cite{1992ApJ...395..130P} in the host galaxy (Table \ref{extinction}).
We do this in the limits that the cooling break is below the optical ($\beta_{\mathrm{opt-X}} = \beta_{\mathrm{X}}= 1.14$) and that the cooling break is far above the optical but below the X-ray region
($\beta_{\mathrm{opt-X}} = \beta_{\mathrm{X}} - 0.5 = 0.64$ below $10^{17}$\,Hz; \citeNP{2004IJMPA..19.2385Z}). 
The R \& I band values of E$_{\mathrm{B-V}}$ are consistent with each other at the $2\sigma$ level in all cases, so one case cannot be favoured over the other.
In the limit where the cooling break is just below the X-ray region, we have lower limits on the host extinction.

Using the host column densities found for the various extinction curves (Table \ref{host_NH}) we calculate the ratio of column density to optical extinction ($N_{\mathrm{H}}$/E$_{\mathrm{B-V}}$, Table \ref{extinction}) in the host and compare to the values of the SMC, LMC and Galaxy \cite{1992ApJ...395..130P}. The expected ratios are 4.5, 2.4 \& 0.48 $\times 10^{22}$\,cm$^{-2}$ respectively, clearly well below the measured values. This is in agreement with previous work \cite{2001ApJ...549L.209G} 
suggesting dust destruction in the circumburst environment.

\subsection{Early optical emission} \label{diss_early}

\citeN{2007ApJ...654L..21S}
suggest that the optical peak at $\sim 540$\,s is the onset of the external shock  -- i.e. the deceleration time, when the jet has swept up enough of the circumburst medium so that the afterglow dominates emission -- however if this was the case we would expect the X-ray to exhibit a peak at the same time. While the X-ray light curve at that time is highly obscured by flares, it does seem that the flares are superimposed on the already decaying X-ray afterglow.

Since early time data has become more common due to the rapid dissemination of burst information from \emph{Swift}, there has been much discussion regarding the reference time, $t_{0}$, for the onset of the power-law decay of the afterglow. Most commonly it is taken as the trigger time $T_{0}$ of the instrument which detected the burst, which is clearly instrument dependent. 
We expect that the light curve will evolve as $F \propto (t - t_{0})^{-\alpha}$, and that it will exhibit a break if it has been assigned the incorrect reference time \cite{2006ApJ...640..402Q}.

To test whether this is a plausible scenario, we fit the optical light curve at times $\gtrsim 540$\,s letting $t_{0}$ as a free parameter. 
The best fit of this suggests that reference time is in fact the trigger time of the burst but since the $\chi^{2}$ function is very flat in that region ($T \ll 500$\,s), a definite minimum is difficult to estimate. 
This is in agreement with the theoretical prediction of \citeN{2007ApJ...655..973K} 
who claim that the reference time should be taken as $T_{0}$. 
The X-ray slope, which is at later times, is unaffected by changes in the reference time much less than the start time of the observations.

If the deceleration time is, as we claim, close to the trigger time, the flat optical light curve from 70\,s -- 310\,s is then likely due to external shock. The blast wave model \cite{2004IJMPA..19.2385Z} in a slow-cooling wind-driven medium predicts a temporal slope $\alpha_{\mathrm{opt}} = 0$, in agreement with the observations ($\alpha_{\mathrm{opt}} = 0.10 \pm  0.10$), for all values of $p$ if $\nu < \nu_{\mathrm{m}} < \nu_{\mathrm{c}}$ where $\nu_{\mathrm{m}}$ is the peak frequency.

We suggest that the rebrightening of $\alpha_{\mathrm{opt}} = -0.9$ in the optical light curve after 310 seconds could  be explained by the change from a wind-driven medium to one of a higher, constant density as detailed by \citeN{2006ApJ...643.1036P}. If, as expected, the progenitor of this burst is a massive star, the circumburst medium is composed of the wind from the star moving into the ISM. This causes a forward shock into the ISM and a reverse shock into the wind, separated by a region of shocked wind with a constant density profile. When the blast wave crosses the reverse shock - shocked wind discontinuity, there is a drop in the cooling frequency. 
This may shift the optical out of the $\nu < \nu_{\mathrm{m}} < \nu_{\mathrm{c}}$ regime into either $\nu < \nu_{\mathrm{c}} < \nu_{\mathrm{m}}$ or $\nu_{\mathrm{c}} < \nu < \nu_{\mathrm{m}}$. 
If this is matched by the standard evolution of the peak frequency, the optical may end up in the $\nu_{\mathrm{m}} < \nu < \nu_{\mathrm{c}}$ or $\nu_{\mathrm{m, c}} < \nu$ regime. The shocked wind has a higher density than the unshocked wind, necessitating a flux increase, causing the observed rebrightening. However, as we shall show in \S \ref{energetics}, the fine tuning of the energetics and circumburst medium parameters make this explanation of the rebrightening less likely.

\subsection{Late afterglow} \label{diss_late}

The X-ray spectral index of the late afterglow, $\beta = 1.14 \pm 0.03$ ($\S$\ref{results_late}), implies an electron power-law index, $p = 2.28 \pm 0.06$, 
assuming that the cooling and peak frequencies are below the X-rays \cite{2004IJMPA..19.2385Z}.  
If the observed X-ray break was a jet break, we would expect the temporal index after the break to be $\alpha_{\mathrm{X}} = p = 2.28$ which is inconsistent with the observed decay of $\alpha = 1.35$. This, in conjunction with the early time of the break ($1.5 ^{+0.4}_{-0.3}$ hr in the rest frame of the GRB), makes a jet break highly unlikely. In the absence of a jet break, the electron power-law index, $p$, implies an X-ray temporal decay index of $\alpha_{\mathrm{X}} = 1.21 \pm 0.05$ in all media, which is not consistent with either of the broken power-law indices (Table \ref{ag_data}). It could however, be in agreement with the single power-law.

\begin{table}[ht]	
  \begin{center} 	
    \caption{Observed late afterglow temporal decays. The possible X-ray break is at $7.4^{+2.1}_{-1.6}$\,hr.} 	
    \label{ag_data} 	
    \begin{tabular}{l l} 
      \hline 
      X-ray (power-law)             & 1.11 $\pm$ 0.02  \\
      X-ray (pre-break)             & 0.89 $\pm$ 0.05  \\
      X-ray (post-break)            & 1.35 $\pm$ 0.06  \\
      Optical                       & 1.154 $\pm$ 0.013  \\
      \hline
    \end{tabular}
\end{center}
\end{table}

Assuming that the optical is in the same frequency regime as the X-rays i.e. above the cooling and peak frequencies, the expected optical temporal decay and spectral indices are predicted to be the same as the X-ray indices, $\alpha_{\mathrm{X}} = \alpha_{\mathrm{opt}} = 1.21$, $\beta_{\mathrm{X}} = \beta_{\mathrm{opt}} = 1.14$. 
This temporal decay index is in agreement with the observed value of $\alpha_{\mathrm{opt}} = 1.15 \pm 0.04$ and consistent with the spectral index assuming host extinction.
If the optical is not in the high-frequency regime, i.e. above either the cooling or peak frequencies, we can rule out a fast cooling regime ($\nu_{\mathrm{c}} < \nu < \nu_{\mathrm{m}}$) as that would imply $\alpha = 0.25$, which is clearly an underestimate. A slow cooling regime ($\nu_{\mathrm{m}} < \nu < \nu_{\mathrm{c}}$) implies a decay of $\alpha = 0.96 \pm 0.05$ in a homogeneous circumburst medium or one of $\alpha = 1.46 \pm 0.05$ in a wind driven medium, each with a spectral index of $\beta = 0.64 \pm 0.06$.

\subsection{Blast wave energetics \& circumburst medium}\label{energetics}

From the measured quantaties we can constrain various physical parameters of the relativistic blast wave and its surroundings; the jet opening angle, the energy of the blast wave, the density and structure of the circumburst medium, the energy in the magnetic field and the relativistic electrons that emit the synchrotron radiation. 
In order to obtain these constraints we use the formulas from \citeN{vdhorst}, while adopting $p=2.28$ and a luminosity distance $d_{\mathrm{L}} = 7.97 \times 10^{28}\,\mathrm{cm}$ 
($\Omega_{\mathrm{M}}=0.27$, $\Omega_{\Lambda}=0.73$, $\mathrm{H}_0=71\,\mathrm{km}\,\mathrm{s}^{-1}\,\mathrm{Mpc}^{-1}$).

Since the temporal behaviour of $\nu_{\mathrm{m}}$, $\nu_{\mathrm{c}}$ and the peak flux $F_{\nu,\mathrm{max}}$ depends on the structure of the circumburst medium, we derive constraints on the physical parameters in two cases: a homogeneous medium, in which the density is constant, and a stellar wind, in which the density as a function of radius is a power-law with index $k = -2$. 
In the homogeneous case $F_{\nu,\mathrm{max}}$ is constant, and $\nu_{\mathrm{m}}$ and $\nu_{\mathrm{c}}$ are decaying power-laws in time, which gives the constraints: $F_{\nu,\mathrm{max}}>F_{\nu_\mathrm{I}}$, $\nu_{\mathrm{m}}<\nu_{\mathrm{I}}$ and $\nu_{\mathrm{c}}<\nu_{\mathrm{I}}$, all at $600$ seconds. 
A fourth constraint comes from the measured I-band flux at 600 seconds, corrected for galactic and host extinction, $F_{\nu_\mathrm{I}}=150\,\mathrm{mJy}$. 
In the stellar wind case three of the four constraints are the same, except for $\nu_{\mathrm{c}}$, because its value increases in time; this means that $\nu_{\mathrm{c}}<\nu_{\mathrm{I}}$ at $7 \times 10^3$ seconds.

The limits on the isotropic equivalent energy we derive are lower limits of $5 \times 10^{54}$\,erg for the homogeneous medium and $2 \times 10^{55}$\,erg for the stellar wind. Since these limits are quite large, we assume that the energies are not much larger. 
This results in values for the fractional energies in electrons and magnetic field of $\varepsilon_{\mathrm{e}}\sim\varepsilon_{\mathrm{B}}\sim 10^{-2}$ (homogeneous) and $\varepsilon_{\mathrm{e}}\sim\varepsilon_{\mathrm{B}}\sim 5 \times 10^{-3}$ (wind). 
The density of the homogeneous medium is then $\sim 10\,\mathrm{cm}^{-3}$, while the wind density parameter $A_{\ast}\sim 1$, i.e. a mass-loss rate of $10^{-5}$ solar masses per year and a wind velocity of $10^3\,\mathrm{km}\,\mathrm{s}^{-1}$. 
With these numbers and the lower limit on the jet-break time of $\sim 10^{6}$\,s we derive a lower limit on the jet opening angle of 8\degr (homogeneous) or 2\degr (wind), resulting in collimation corrected energies of $5 \times 10^{52}$ and $1 \times 10^{52}\,\mathrm{erg}$ respectively. These energies are quite large, but not impossible, especially if you consider that the early optical emission could be due to sustained energy injection by the progenitor, providing even more energy for the blast wave than there was for the prompt emission. Furthermore, we note that a blast wave at this redshift and suffering significant host extinction has to be very energetic to produce the observed fluxes.

We have suggested that the early optical behaviour could be explained by the transition from a massive stellar wind to a homogeneous medium. To get the correct temporal slopes this implies that $\nu_{\mathrm{I}}<\nu_{\mathrm{m}}<\nu_{\mathrm{c}}$ from 70 -- 310\,s, and $\nu_{\mathrm{m,c}}<\nu_{\mathrm{I}}$ after 540\,s. This requires some fine-tuning of the parameters, in particular $\nu_{\mathrm{c}}$ has to decrease significantly because of the density jump, and at the same time $\nu_{\mathrm{m}}$ has to pass through the I band, since the peak frequency does not depend on the density and only mildly on the density structure. 
The isotropic equivalent energy that one derives in this case is even higher than before, namely $6 \times 10^{55}\,\mathrm{erg}$, while $\varepsilon_{\mathrm{e}}\sim 4 \times 10^{-3}$ and $\varepsilon_{\mathrm{B}}\sim 3 \times 10^{-3}$. The density of the homogeneous medium is again $\sim 10\,\mathrm{cm}^{-3}$, but $A_{\ast}\sim 0.15$ in this case. From these parameters we get a lower limit for the jet opening angle of 6\degr and a collimation corrected energy of $3 \times 10^{53}\,\mathrm{erg}$. This value for the energy is very high and difficult to accommodate within current progenitor models, making the transition from a wind to a homogeneous medium a less likely explanation for the early optical emission.

We note that with these physical parameters the synchrotron self-absorption frequency, $\nu_{\mathrm{a}}$, lies well above 8.64\,GHz at 3.95 days when the $3\sigma$ limit was obtained by \citeN{2006GCN..4761....1F}. The radio limit is hence consistent with both a homogeneous and wind-driven circumburst medium.

\subsection{Energy injection}\label{inject}

Although the X-ray light curve is best fit with a broken power-law, it is difficult to accommodate the temporal slopes within our current afterglow models. The pre-break slope is shallower than the simultaneous optical slope, which is only possible if the circumburst medium is a stellar wind and $\nu_{\mathrm{m}} < \nu_{\mathrm{opt}} < \nu_{\mathrm{c}} < \nu_{\mathrm{X}}$. This would imply that the optical slope should be $\alpha_{\mathrm{opt}} = 1.46$ and that the X-ray slope $\alpha_{\mathrm{X}} = 1.21$, which is much higher than the observed pre-break values. 
Continued energy injection from the central engine \cite{2006ApJ...642..389N} 
would cause this slope to be shallower for the period of injection before reverting to the original value. For the optical temporal slope this means that the energy has to increase as $t^{0.37 \pm 0.07}$, while the pre-break X-ray slope indicates $t^{0.30 \pm 0.08}$. The break in the X-ray light curve at $\sim 7.4\,\mathrm{hr}$ is then the end of the energy injection phase. The post-break X-ray temporal slope is consistent with the spectral slope at a $2\sigma$ level for $\nu_{\mathrm{m,c}} < \nu_{\mathrm{X}}$. The temporal break can not be due to the passage of $\nu_{\mathrm{c}}$, since there is no spectral change observed between the pre-break and post-break spectra. 

From the constraints on the ordering of the peak, cooling and observing frequencies we can again obtain limits on the physical parameters of the blast wave and its surroundings. For the calculations we adopt that the isotropic equivalent energy is proportional to $t^{0.34}$, changing the temporal scaling laws of the peak flux, peak frequency and cooling frequency into $F_{\nu,\mathrm{max}}\,\propto\,t^{-0.33}$, $\nu_{\mathrm{m}}\,\propto\,t^{0.67}$ and $\nu_{\mathrm{c}}\,\propto\,t^{-1.33}$. 
The values for the physical parameters we derive are quite similar to the values in \S \ref{energetics}: $\varepsilon_{\mathrm{B}}\sim 10^{-3}$, $\varepsilon_{\mathrm{e}}\sim 10^{-2}$, $A_{\ast}\sim 1$, the isotropic equivalent energy at the end of the energy injection phase $\sim 5 \times 10^{55}\,\mathrm{erg}$, the jet-opening angle $> 2\degr$, and the collimation corrected energy $\sim 2 \times 10^{52}\,\mathrm{erg}$. So again, while the energy requirements are at the high end of the distribution, they are not unprecedented.


\section{Conclusion}

We have analysed the optical, X-ray and $\gamma$-ray data of  GRB\,060210 from the time of the prompt emission up until the X-ray afterglow was no longer detectable by \emph{Swift}-XRT.

The early flaring X-ray \& $\gamma$-ray emission implies internal shocks, indicative of prolonged engine activity. The light curves of the two regimes match well, even displaying a common flare at $\sim 100$\,s. The agreement of the X-ray flares to the curvature effect and the clearly demonstrated spectral evolution support the claim that the flares are indeed due to internal shocks. 
The simultaneous optical emission, on the other hand, is consistent with an external shock afterglow with a frequency below that of both the peak \& cooling frequencies. We have shown that such an external shock's deceleration time is close to the trigger time and should hence dominate the optical emission at the time of our observations.

The late afterglow ($\gtrsim 600$\,s) may be explained by either of two models with an electron power-law index, $p = 2.28$, derived from the X-ray spectral index, $\beta_{\mathrm{X}}$: 
a single temporal power-law in both optical and X-rays with both observing bands above the cooling frequency; 
or a broken power-law in X-rays in which optical and pre-break X-ray temporal slopes can be explained by continued activity of the central engine. In the latter case the cooling frequency is between the optical and X-ray frequencies and the observed X-ray temporal break indicates the termination of the central engine activity. Though both are consistent with the temporal and spectral slopes, we favour the continued central engine activity since the $\chi^{2}$ of the single power-law fit is significantly worse. The collimation corrected energy requirements of $\sim 10^{52}\,\mathrm{erg}$ in both cases, are at the high end of the distribution for GRBs but are certainly not unprecedented.

Comparing the column density in the host to the optical extinction we find a higher than expected value which supports the notion that GRBs can cause destruction of dust in the circumburst environment.

\begin{acknowledgements}
PAC acknowledges the support of NWO under grant 639.043.302 and of the University of Leicester SPARTAN exchange visit programme, funded by the European Union Framework 6 Marie Curie Actions. 
APB, CPH, PTOB, KLP, ER and RW gratefully acknowledge support from PPARC.
We thank R. Starling for useful discussions regarding host extinction.
The Faulkes Telescopes are operated by the Las Cumbres Observatory Global Telescope Network.
The authors acknowledge benefits from collaboration within the EU FP5 Research Training Network ``Gamma-Ray Bursts: An Enigma and a Tool" (HPRN-CT-2002-00294). 
This research has made use of data obtained through the High Energy Astrophysics Science Archive Research Center Online Service, provided by the NASA/Goddard Space Flight Center. 
\end{acknowledgements}



\end{document}